\documentclass[twocolumn,prl,showpacs,preprintnumbers,amsmath,amssymb,floatfix]{revtex4}
\pdfoutput=1
\usepackage{graphicx}
\usepackage{dcolumn}
\usepackage{bm}
\begin{document}
\preprint{PRL}
\title{Paired atom laser beams created via four-wave mixing}
\author{R.G. Dall}%
\author{L.J. Byron}%
\author{A.G. Truscott}
\email{andrew.truscott@anu.edu.au}
\affiliation{ARC Centre of Excellence for Quantum-Atom Optics}
\affiliation{Research School of Physical Sciences and Engineering, 
Australian National University, Canberra, ACT 0200, Australia.}%
\author{G.R. Dennis}%
\author{M.T. Johnsson}%
\author{J.J. Hope}
\affiliation{ARC Centre of Excellence for Quantum-Atom Optics}
\affiliation{Department of Quantum Physics, Australian National University, Canberra, ACT 0200, Australia.}%
\date{\today}
\begin{abstract}
A method to create paired atom laser beams from a metastable helium atom laser via four-wave mixing is demonstrated.  Radio frequency outcoupling is used to extract atoms from a Bose Einstein condensate near the center of the condensate and initiate scattering between trapped and untrapped atoms.  The unequal strengths of the interactions for different internal states allows an energy-momentum resonance which leads to the creation of pairs of atoms scattered from the zero-velocity condensate.  The resulting scattered beams are well separated from the main atom laser in the 2-dimensional transverse atom laser profile.  Numerical simulations of the system are in good agreement with the observed atom laser spatial profiles, and indicate that the scattered beams are generated by a four-wave mixing process, suggesting that the beams are correlated.
\end{abstract}
\maketitle
Sources of matter waves gained a dramatic improvement with the achievement of Bose-Einstein condensation (BEC) in dilute gases and the development of the atom laser \cite{BEC1,KetterleAL}.  Like optical lasers before them, atom lasers can produce Heisenberg-limited beam profiles \cite{Esslinger,Aspect} and promise high spectral density through their dramatically lower linewidth \cite{WisemanDefiningThe}.  Another exciting possibility resulting from having such a coherent source of atoms is the generation of non-classical matter waves through entangled beams.  Such entangled beams are useful for tests of quantum mechanics, and are required to  perform Heisenberg-limited interferometry \cite{Dowling98,Reid88}.  In this Letter, we show that the asymmetric scattering rates between internal states of metastable helium (He*) cause well-defined peaks in the output of an atom laser.  These peaks are due to a four-wave mixing (FWM) process, and are experimentally demonstrated.

A nonlinear process is required to produce entanglement, and one of the advantages of atomic systems over optical systems is that there are strong inherent nonlinearities due to the atomic interactions, although these interactions can also lead to complications.  These nonlinearities allow certain analogs of nonlinear optical experiments such as FWM and Kerr squeezing to be performed directly in the atomic sample \cite{KerrReference}.  Both of these produce entanglement in optical systems.
FWM in a trapped BEC has been demonstrated experimentally in configurations where three distinct momentum states generated a fourth \cite{PhillipsFWM}, and where two momentum states generated a pair of correlated atomic beams \cite{KetterleFWM}.  These experiments demonstrated that the output was phase coherent, but the correlation properties were not measured.  More recently the pair correlations in a spontaneous scattering of two colliding condensates were measured using the single atom detectors available for He* atoms \cite{WestbrookFWM}.  

Using these existing sources of entangled pairs of atoms for interferometric experiments will be complicated by the high densities of the sources, where the nonlinearities that generated the correlations ultimately degrade the long term coherence of the sample.  While recent experiments have increased the coherence of atom interferometers by several orders of magnitude by reducing the nonlinearities with a Feshbach resonance~\cite{Fattori2008,Gustavsson2008}, this precludes the production of entangled pairs. In our scheme the nonlinear interactions are used to drive FWM in the magnetically trapped condensate, but the resulting untrapped beams that propagate in free space are dilute, potentially avoiding the decoherence problem. Using atoms in the untrapped state also makes the beams insensitive to magnetic field inhomogeneities.  We show that pairs of beams can be produced simply by the process of radio frequency (RF) outcoupling from a He* BEC, without the need for Feshbach resonances, optical traps or scattering pulses.  Unlike the previous methods, which required pairs of atoms travelling at high kinetic energies as a source, this process involves scattering between atoms initally in the same  zero momentum state to create states with non-zero momentum.  The energy-momentum resonance comes from the mean field conditions that are obtained during outcoupling from the condensate.  Semiclassical and field theoretic simulations of the experiment show that the beams are generated by the same FWM process that generated entangled atom pairs in the earlier experiments.

Our experimental setup for creating a He* BEC has been reported elsewhere \cite{dall06}.  Briefly, we use a cryogenic beamline to produce a $\text{He}^*$ magneto-optic trap (MOT).  Atoms are extracted from this low vacuum MOT into a high vacuum MOT via an LVIS$^{+}$ setup. Atoms are transferred from the MOT into a BiQUIC magnetic trap (with axial trapping frequency of 55 Hz and radial trapping frequency of 1020 Hz), where a BEC is produced using forced RF evaporative cooling. Using this setup we are able to create almost pure BECs containing up to $5\times10^6$ atoms.  To produce our atom laser beam, we use RF photons to spin flip the BEC atoms from the $m=1$ magnetically trapped state to the $m=0$ untrapped state. 

After outcoupling, atoms in the atom laser beam fall under gravity for a distance of $4\, \text{cm}$ until they strike a double stacked multi-channel plate (MCP).
We image the phosphor screen with a CCD camera with a resolution of approximately $150\, \mu\text{m}$ at the MCP.  To remove any nonuniformities caused by spatial variations in the gain of our MCP, we divide all our images by a flat field image, produced by dropping atoms from a MOT onto our detector.  Since the MOT temperature is of order $\sim1\, \text{mK}$, the spatial profile of the MOT uniformly illuminates our MCP.  Although $m=-1$ atoms are produced by the outcoupling, especially for high RF powers, they are in general accelerated away from our detector by the magnetic trap field.  Those that are accelerated towards the detector do not show up in our images since they arrive much earlier than the CCD trigger time.

\begin{figure}
        \begin{center}
        \includegraphics[width=86mm]{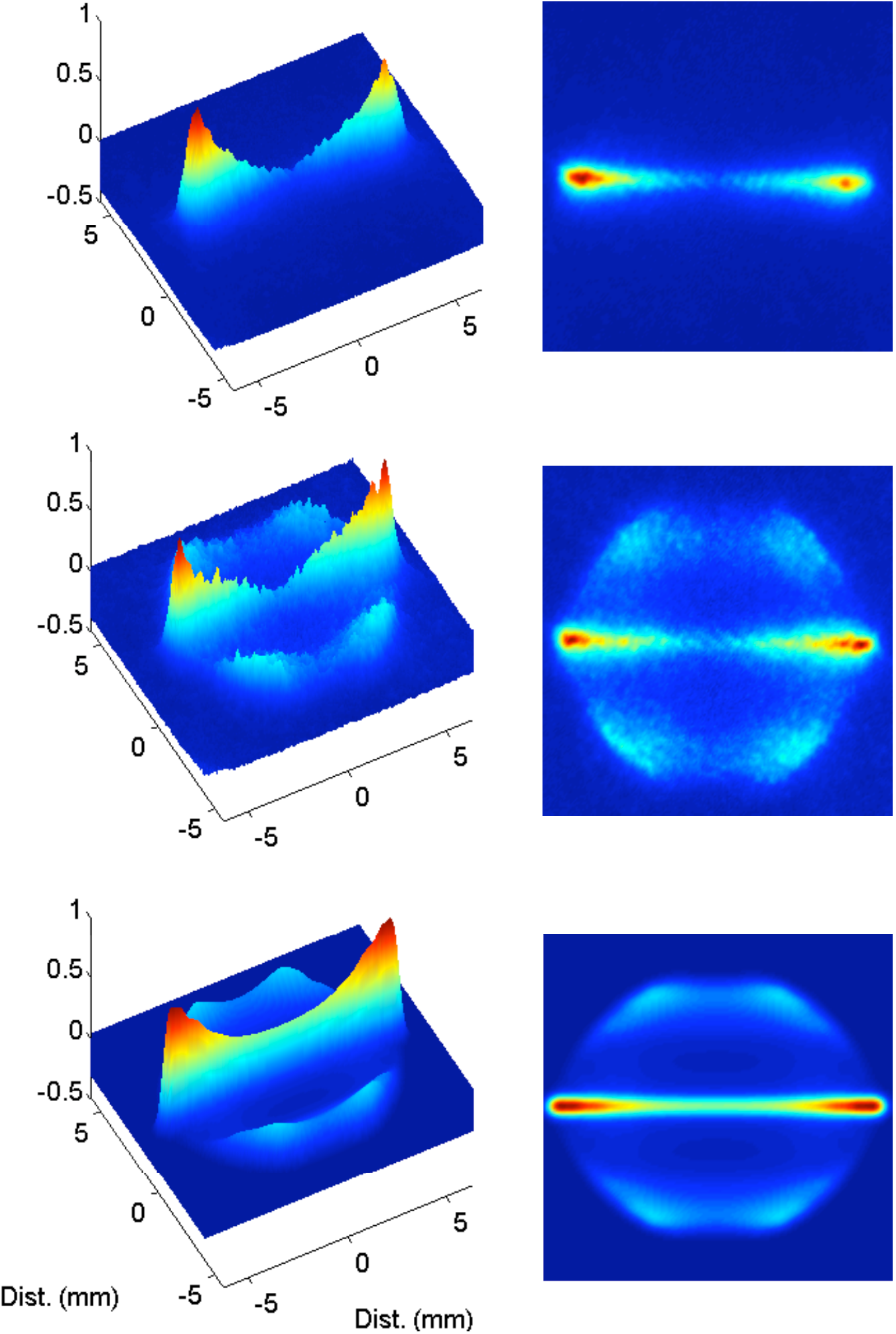}
        \end{center}
\caption{(color online) First two rows show experimental atom laser spatial profiles observed on our MCP 4cm below the trap, in a 3-D rendering (left) and the 2-D image (right).  Both sets of data were taken for an outcoupling detuning of $2\, \text{kHz}$, however the Rabi frequency is increased by an order of magnitude between the two sets.  The upper row shows the usual $\text{He}^*$ atom laser~\cite{dall2007}, while the middle row demonstrates the appearance of the resonant scattering peaks.  The bottom row is the result of a simulation of the second experiment.  The weak axis of the trap is aligned along the vertical axis of the right hand images.}
\label{figExperimentalPeaks}
\end{figure}

Figure~\ref{figExperimentalPeaks} shows the dramatic change in the atom laser spatial profile when the condensate undergoes the resonant FWM process. In the case of low outcoupling Rabi frequencies (upper row) the FWM process does not occur and we see the usual double peaked $\text{He}^*$ atom laser profile~\cite{dall2007}.  When the Rabi frequencies are high enough to initiate the FWM process (middle row) the atoms in the atom laser beam are scattered to form a halo around the main atom laser beam.  Due to conservation of energy the outer diameter of this halo corresponds to a maximum kinetic energy given by the chemical potential.  As well as the ring structure, four peaks are observed on the outskirts of the profile. Each peak contains 11--14\% of the total outcoupled atoms, in agreement with the value of 11\% obtained from the simulation results shown in the lower row of Fig.~\ref{figExperimentalPeaks}. 

\begin{figure}
\begin{center}
\includegraphics[width=86mm]{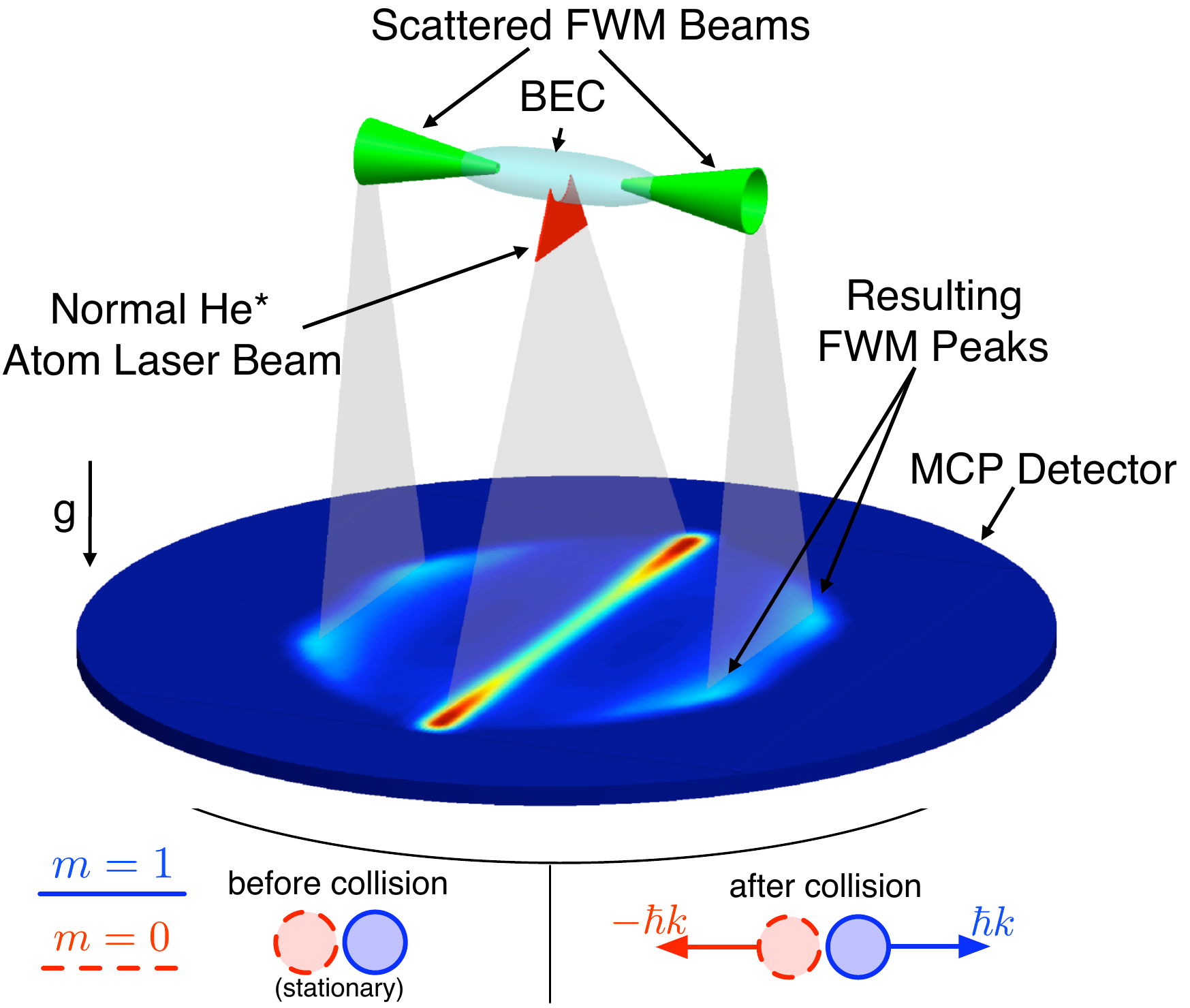}
\caption{(color online) Schematic diagram of experimental configuration. The four-wave mixing process generates momentum components along the weak trapping axis, which get outcoupled to two cones. After outcoupling from the condensate, atoms fall $4\, \text{cm}$ to the MCP detector.\label{fig:Lesa}}
\end{center}
\end{figure}

These peaks arise from two momentum-correlated cones of particles scattering out of the condensate and falling to the detector, as shown in Fig.~\ref{fig:Lesa}. The cones themselves are generated via a FWM process that populates momentum modes lying along the weak trapping axis, which expand to rings due to the mean field repulsion in the tight trapping direction as they leave the BEC. Atoms in these cones then fall under gravity onto the detector, where the time-integrated flux is converted to a spatial density distribution, and each momentum ring appears as a double peak. The background halo is produced during the initial switch-on of the atom laser during which the peaks produced by the FWM process sweep in position from the main atom laser profile towards their final position in Fig.~\ref{figExperimentalPeaks}.

In order to understand how these peaks are produced we carried out a detailed numerical simulation of the experiment. The master equation we used to describe the experiment is given by
\begin{equation}
\frac{\partial \hat{\rho}}{\partial t} = \frac{-i}{\hbar} \left[\hat{H}, \hat{\rho}\right] + \frac{54}{5} K^{\text{(unpol)}}_{^4\text{He}} \int \mathcal{D}\left[\hat{\Psi}^{\text{(mol)}}_{J=0}(\mathbf{x})\right] \hat{\rho}\, d\mathbf{x}
\label{eq:MasterEquation}
\end{equation}
where
\begin{align}
    \begin{split}
    \hat{H} = &\textstyle \sum_j \int \hat{\Psi}_j^\dagger(\mathbf{x})\left(-\frac{\hbar^2 \nabla^2}{2 M} + j V_\text{trap}(\mathbf{x}) \right) \hat{\Psi}_j(\mathbf{x})\, \mathrm{d}\mathbf{x} \\
              &\textstyle + \sum_{i j} \int \hat{\Psi}_i^\dagger(\mathbf{x}) \sqrt{2} (\delta_{i,j+1} + \delta_{i, j-1}) \hbar \Omega \hat{\Psi}_j(\mathbf{x})\, \mathrm{d}\mathbf{x} \\
              &\textstyle + \sum_{i j} \frac{U_{i j}}{2} \int \hat{\Psi}_i^\dagger (\mathbf{x}) \hat{\Psi}_j^\dagger (\mathbf{x}) \hat{\Psi}_j(\mathbf{x}) \hat{\Psi}_i(\mathbf{x})\, \mathrm{d}\mathbf{x}, 
    \end{split} \label{eq:Hamiltonian}
\end{align}
and where $\mathcal{D}[\hat{c}]\hat{\rho} \equiv \hat{c}\hat{\rho}\hat{c}^\dagger - \frac{1}{2} \left(\hat{c}^\dagger \hat{c} \hat{\rho} + \hat{\rho}\hat{c}^\dagger \hat{c} \right)$, $\hat{\Psi}_m$ are the field operators for the magnetic sublevels $m$, $\Omega = 350\, \text{Hz}$ is the Rabi frequency of the RF outcoupling, $V_\text{trap}(\mathbf{x}) = \frac{1}{2} M \left(\omega_x^2 x^2 +\omega_y^2 y^2+ \omega_z^2 z^2\right)$ is the trapping potential, $\displaystyle U_{ij} = 4 \pi \hbar^2 a_{i,j}/M$ is the nonlinear interaction strength, $a_{i, j}$ is the s-wave scattering length between the magnetic sublevels $i$ and $j$, and $K^\text{(unpol)}_{^4\text{He}}$ is the Penning ionisation rate for an unpolarised thermal sample of $\text{He}^*$ ~\cite{stas06}.  The scattering lengths for $\text{He}^*$ are $a_{1,-1} = 3.33\, \text{nm}$, $a_{0,0} = 5.56\, \text{nm}$ and $7.51\, \text{nm}$ for all other combinations of internal states \cite{leo2001}.  The Penning ionisation loss term in Eq.~(\ref{eq:MasterEquation}) was obtained by assuming a position-independent loss for the molecular state with total angular momentum of 0, $\hat{\Psi}^{\text{(mol)}}_{J=0}(\mathbf{x}) = \frac{1}{\sqrt{6}} \left( 2 \hat{\Psi}_1(\mathbf{x}) \hat{\Psi}_{-1}(\mathbf{x}) - \hat{\Psi}_0(\mathbf{x}) \hat{\Psi}_0(\mathbf{x}) \right)$, which is the dominant contribution to Penning ionisation at condensate temperatures~\cite{stas06}.  The rate constant was determined by matching the ionisation rate for a thermal sample of $\text{He}^*$ to the results reported in Ref.~\cite{stas06}.

The important dynamics of this system occur inside or near the condensate, and the evolution far below this region is well described by free-fall.  We therefore employ a two-step method where we first calculate the evolution of the three magnetic substates in a restricted region enclosing the condensate and then propagate the momentum flux density at the edge of the simulation region classically to determine the profile on the detector located $4\, \text{cm}$ below the BEC. To take into account the estimated expansion due to mean-field repulsion in the beam itself we use an approximate method where we convolve the detector profiles with a Gaussian of width of $200\, \mu\text{m}$.


The results of these simulations are shown in the lower row of Fig.~\ref{figExperimentalPeaks}, and agree well with the experimental results in the middle row. As borne out by these simulations, the process forming the peaks is the scattering of stationary trapped and untrapped atoms, which produce pairs of atoms with oppositely-directed momenta in the weak trapping dimension. This is different to the usual FWM process where two atoms with nonzero kinetic energy collide, conserving kinetic energy to scatter into different momentum modes.

To determine the source of the additional kinetic energy, we consider that the scattered atoms will cause interference with any background atoms.  This interference causes density gratings for both of the internal states.  In the case of equal scattering lengths the interaction energy is unchanged, as the total density is unchanged. 
However, $\text{He}^*$  has a reduced scattering length for 0--0 collisions as compared to 0--1 and 1--1 collisions, so the interaction energy is \emph{reduced} by the formation of the gratings.  In the case of equal, uniform background densities in both states, the change in the interaction energy is
\begin{align}
    \Delta \big<\hat{H}_\text{int} \big> = & U (1 - \kappa) \int \left( 2 \left| \Psi \right|^2 \varepsilon \cos(k x) - \varepsilon^2 \cos^2(k x)\right) \mathrm{d}\mathbf{x},
\end{align}
where $\left|\Psi\right|^2$ is the background density of both atomic fields,  $\varepsilon$ is the amplitude of the density gratings in both states, $U = U_{11} = U_{10}$ and $\kappa = a_{0,0}/a_{1,1} < 1$ is the ratio of the scattering lengths such that $\kappa U = U_{00}$.  The first term in this expression averages to zero over one period of the density grating and the second term is always negative, indicating that the interaction energy has \emph{reduced} due to the scattering. Thus the increased kinetic energy after scattering is balanced by the reduced mean field energy, allowing the scattering to take place, even though the atoms were initially at rest. A resonance between this reduced interaction energy and the increased kinetic energy of the scattered atoms causes well-defined beams to be produced by FWM.

\begin{figure}
        \begin{center}
        \includegraphics[width=76mm]{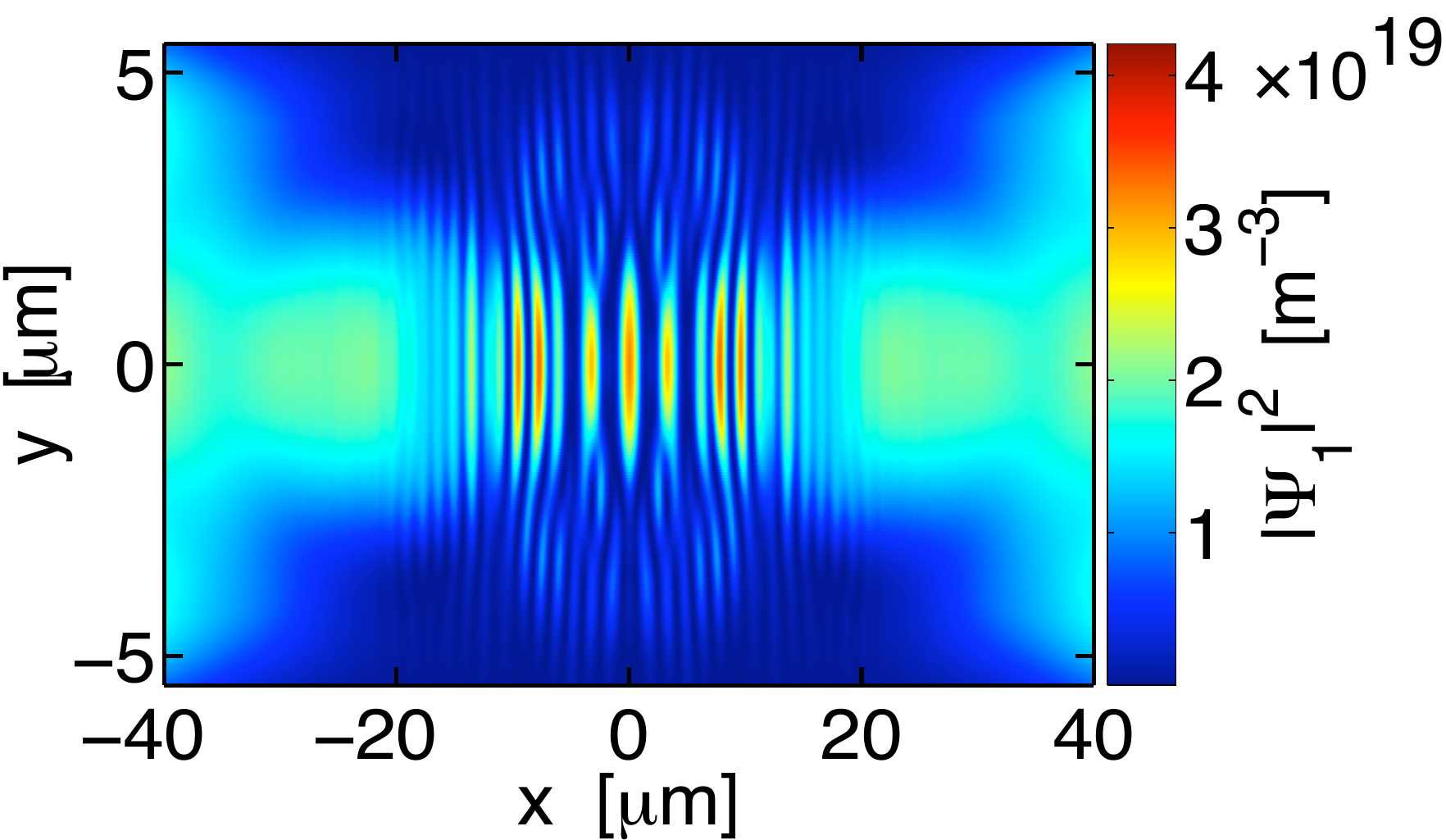}
        \end{center}
\caption{(color online) A Gross-Pitaevskii simulation of the in-trap condensate density after $3.6\, \text{ms}$ of outcoupling, magnified to show the central region of the condensate, where the scattering-induced grating has high visibility. The weak and tight trap directions are $x$ and $y$ respectively.}
\label{fig:combinedtheoryexp1}
\end{figure}

Generalising this argument to allow different density backgrounds for the $m=1$ and $m=0$ states yields the wavenumber corresponding to the energy-momentum resonance of the scattered atoms as
\begin{align}
k &= \frac{1}{\hbar} \sqrt{2 M U \left( 2 \left| \Psi_1 \right| \left| \Psi_0 \right| - \left| \Psi_1 \right|^2 - \kappa \left| \Psi_0 \right|^2 \right)}. \label{eq:ResonantMomentum}
\end{align}
Figure~\ref{fig:combinedtheoryexp1} shows the result of a simulation of the density of the BEC after a period of outcoupling showing the appearance of this density grating in the trapped BEC.

The FWM process discussed in this Letter can be either stimulated by the presence of an existing atomic population in the final momentum states, or seeded spontaneously if those target states are initially empty.

The stimulated process occurs if the outcoupling surface is chosen to be an ellipsoidal shell, for example by detuning the outcoupling from the center of the condensate by $2\, \text{kHz}$. This causes a range of densities in the trapped and untrapped fields to exist over the surface. From Eq.~(\ref{eq:ResonantMomentum}) we see that this corresponds to a range of momenta that are potentially resonant, from zero up to some maximum value. Atoms accelerated in the mean-field and trap potentials are amplified by this FWM scattering, allowing resonant stimulated scattering into higher and higher final momentum modes.  This sweeps out the halo shown in Fig.~\ref{figExperimentalPeaks}, and finally reaches the momentum corresponding to the peaks which then become heavily populated.

Spontaneous processes cannot be described by a mean-field model, so we simulated this system using the stochastic truncated Wigner (TW) method \cite{truncatedWigner}.  As the detuning is reduced below $2\, \text{kHz}$, the outcoupling surface decreases in size and the minimum resonant momentum $\hbar k$ for the FWM process increases while the maximum resonant momentum remains approximately constant. In this situation the lower momentum modes near zero are no longer resonant, and the FWM process can only be started spontaneously. After the spontaneous scattering has taken place, these momentum modes then build up over time through stimulated scattering.  At these lower detunings, in the TW simulations the extra peaks are better defined due to the narrower range of resonant momenta (though weaker and more difficult to see experimentally).  Importantly, the peaks are \emph{non-existent} using semiclassical simulations, where spontaneous scattering does not occur. Moreover, the resonant momentum for the gratings of $1.2\times 10^6\, \text{m}^{-1}$ is in excellent agreement with the result given by Eq.~(\ref{eq:ResonantMomentum}) of $1.3\times 10^6\, \text{m}^{-1}$.  This demonstrates that the momentum modes are generated by a spontaneously-seeded four-wave mixing process, where two atoms in the condensate with near-zero momentum are scattered into modes with opposite momentum along the weak trapping axis. This process will produce EPR entanglement between the trapped and untrapped scattered atoms. This may lead to entanglement between atoms on opposing points of the cones in Fig.~\ref{fig:Lesa}, but due to the output dynamics and the vertical integration of the detection scheme, this would be difficult to measure. However, number difference squeezing between the opposing peaks should be far more robust.

As the maximum wavenumber for these resonances in both the spontaneously and coherently seeded process corresponds to a wavelength larger than the Thomas-Fermi radius in the tight trapping dimension, the energy and momentum resonances only co-exist in the weak trapping dimension, leading to the asymmetry in Fig.~\ref{fig:Lesa}. If the density in the untrapped state is too low, no grating will form as the wavenumber for the energy-momentum resonance given by Eq.~(\ref{eq:ResonantMomentum}) will be imaginary as is the case for low outcoupling (upper row Fig.~\ref{figExperimentalPeaks}).

We have shown that appropriate outcoupling from a $\text{He}^*$ BEC can produce well-defined additional peaks in the output beam.  Field theoretic and semiclassical models show that these peaks are formed from scattering of pairs of atoms in BEC, and are therefore entangled upon formation.  It remains to be seen whether useful entanglement remains after the outcoupling process.  These beams are produced directly via RF outcoupling in a magnetic trap. The potential advantages of these correlated beams are that they are spatially well separated from the background of the atom laser and that the quasi-continuous dilute beam will likely remain coherent over larger timescales than trapped fields.  

\begin{acknowledgments}The authors wish to acknowledge the technical assistance of Stephen Battisson in the design and construction of the He* beamline, and Ken Baldwin, John Close and Craig Savage for their suggestions. This work is supported by the Australian Research Council Centre of Excellence for Quantum-Atom Optics and the APAC National Supercomputing Facility.
\end{acknowledgments}

\end{document}